\documentclass[conference]{IEEEtran}
\IEEEoverridecommandlockouts
\usepackage{cite}
\usepackage{amsmath,amssymb,amsfonts}
\usepackage{algorithmic}
\usepackage{graphicx}
\usepackage{textcomp}
\usepackage{xcolor}
\usepackage[inline]{enumitem}
\usepackage{array}

\usepackage[caption=false]{subfig}

\def\BibTeX{{\rm B\kern-.05em{\sc i\kern-.025em b}\kern-.08em
    T\kern-.1667em\lower.7ex\hbox{E}\kern-.125emX}}

\begin{document}

\title{A Cyberpunk 2077 perspective on the prediction and understanding of future technology \\
\thanks{Cyberpunk 2077 screenshots used with permission. CD PROJEKT RED is not involved in this project and is not an official partner in any capacity.}
}

\author{
\IEEEauthorblockN{Miguel Bordallo López$^{\star}$$^{\dagger}$, Constantino Álvarez Casado$^{\star}$}
\IEEEauthorblockA{\textit{$^{\star}$Center for Machine Vision and Signal Analysis, University of Oulu} \\
\textit{$^{\dagger}$VTT Technical Research Centre of Finland}\\
Oulu, Finland \\
\{miguel.bordallo, constantino.alvarezcasado\}@oulu.fi}

}  

\maketitle

\begin{abstract}

Science fiction and video games have long served as valuable tools for envisioning and inspiring future technological advancements. This position paper investigates the potential of Cyberpunk 2077, a popular science fiction video game, to shed light on the future of technology, particularly in the areas of artificial intelligence, edge computing, augmented humans, and biotechnology. By analyzing the game's portrayal of these technologies and their implications, we aim to understand the possibilities and challenges that lie ahead. We discuss key themes such as neurolink and brain-computer interfaces, multimodal recording systems, virtual and simulated reality, digital representation of the physical world, augmented and AI-based home appliances, smart clothing, and autonomous vehicles. The paper highlights the importance of designing technologies that can coexist with existing preferences and systems, considering the uneven adoption of new technologies. Through this exploration, we emphasize the potential of science fiction and video games like Cyberpunk 2077 as tools for guiding future technological advancements and shaping public perception of emerging innovations.

\end{abstract}


\section{Introduction}

The rapid pace of technological advancements has led to an ever-increasing interest in understanding and predicting the future of technology. The capacity of science fiction to stimulate actual scientific exploration, technological advancement and their impact on society remains a topic of sustained fascination \cite{landon2014science,gunn2002road}, as seen in works such as \textit{Neuromancer} \cite{gibson1984neuromancer}, \textit{Snow Crash} \cite{stephenson1992snowcrash}, \textit{Do Androids Dream of Electric Sheep?} \cite{dick1968do}, \textit{Little Brother} \cite{doctorow2008little},  \textit{Star Trek} \cite{Roddenberry1966startrekseries} and \textit{The Three-Body Problem} \cite{liu2015three}. Film serves as a compelling medium for illustrating the intricate relationship between science fiction and technological progress. These cinematic depictions enrich the ongoing scholarly discourse on the ethical and technological implications of emerging innovations, particularly in sectors like healthcare and aerospace \cite{samuel2019can}. More recently, video games have emerged as an interactive medium that offers unique opportunities to explore futuristic concepts and ideas, with some experts arguing that these ideas are inevitable \cite{kelly2016inevitable}.

One such video game, Cyberpunk 2077, has garnered attention for its portrayal of a technologically advanced society and the potential challenges and opportunities it presents. The game's developers have detailed the artistic and technological aspects of creating its immersive cyberspace \cite{ankermann2021tech}. Its companion book, The World of Cyberpunk 2077 \cite{batylda2020world}, offers an in-depth look into the game's lore and setting and provides a comprehensive understanding of the game's world.  

Set in a dystopian future, Cyberpunk 2077 presents a detailed vision of an urban landscape dominated by powerful corporations that shape and govern the direction of technological advancements. This backdrop vividly intertwines social, political, and economic themes, providing a rich narrative for players to navigate. Cyberpunk 2077 has not only captured the imagination of gamers worldwide but has also sparked discussions among academics, researchers, and industry professionals about the likelihood and implications of such a future \cite{lay2020cyberpunk}\cite{ bostrom2014superintelligence}\cite{fox2021possible}\cite{specialissue}.
\begin{figure}[b!]
  \begin{center}
  \vspace{-5mm}
    \includegraphics*[width=0.48\textwidth]{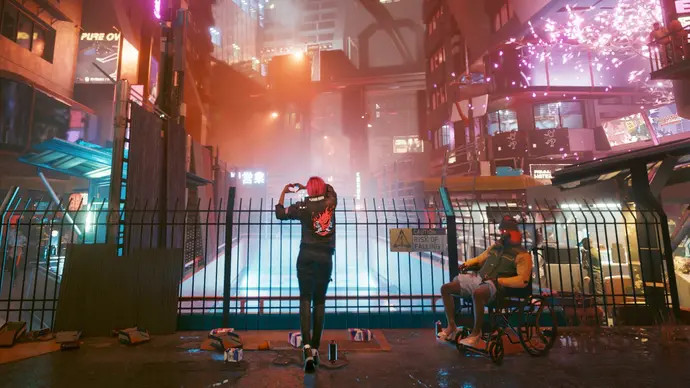}
  \end{center}
  \vspace{-3mm}
    \caption{A screenshot of Cyberpunk 2077, showing details of a urban landscape and society.}
  \label{fig:cp_main}
  \vspace{-3mm}
\end{figure}
The game presents a nuanced view of technology, exploring themes such as disability, gender, and the relationship between humans and machines \cite{banbury2022so}\cite{steele2023chippin}. These themes raise significant ethical and social questions, illustrating the potential of video games as a medium for stimulating debate and influencing future advancements. As an illustration, Fox \cite{fox2021possible} delves into the portrayal of disability in Cyberpunk 2077 and the potential trajectory of cyborg healthcare. Notably, while the game offers detailed depictions of technologically advanced societies, history underscores that the uptake of innovations is often uneven. This disparity is shaped by factors like economic differences, cultural inclinations, and regulatory landscapes \cite{jameson2002uneven}\cite{rogers2010diffusion}\cite{norris2001digital}, a notion echoed profoundly in the game \cite{batylda2020world}.

This article aims to examine the key technological themes and advancements presented in Cyberpunk 2077 and compare them with the current state of technology. By doing so, we hope to identify existing challenges and opportunities for further development in various domains, including human augmentation, brain-computer interfaces, simulated reality, digital representation, smart homes and appliances, autonomous vehicles, and advanced smart assistants. Furthermore, we will discuss the importance of designing technologies that can coexist with existing preferences and technologies, ensuring a seamless transition towards a more technologically advanced society \cite{kelly2016inevitable}\cite{fox2021possible}.

\section{Literature Review}
In this section, we review existing literature on the role of science fiction in predicting and inspiring technological advancements, as well as studies on the specific technologies mentioned in the presentation. Our aim is to establish the context and show how the paper's ideas fit into the broader academic conversation.

\subsection{Science Fiction and Technological Advancements}
Science fiction has a long history of providing a platform for speculating about future technological innovations and their impact on society. The genre often serves as a source of inspiration for researchers and technologists, with numerous real-life inventions tracing their origins to ideas first presented in science fiction stories \cite{turing1950computing}\cite{bassett2013better}. For instance, the concept of the internet was presaged by authors such as William Gibson in his novel \textit{Neuromancer} \cite{gibson1984neuromancer}. The 1966 television series \textit{Star Trek} presciently introduced the concept of mobile communication devices, a vision that materialized with Motorola's launch of the inaugural flip phone, the \textit{StarTAC}, in 1996 \cite{Bennett2017StarTrekReviewTech}\cite{atkin1995science}\cite{jordan2018fiction}. Similarly, the idea of a tablet computer can be traced back to Arthur C. Clarke's \textit{2001: A Space Odyssey} \cite{Clarke1968}. Stanley Kubrick's seminal 1968 work, provided also an early examination of the ethical and computational complexities inherent in artificial intelligence, exemplified by the character HAL 9000 in the subsequent same named film \cite{kubrick19682001}. Similarly, George Lucas's \textit{Star Wars} series initiated the concept of 3D holographic technology, an idea that has subsequently transitioned into practical applications \cite{lucas1977star}. The 1995 film \textit{Johnny Mnemonic} notably anticipated contemporary virtual reality technologies, including VR headsets, hand controllers, and real-time hand tracking within three-dimensional environments \cite{gibson1986Mnemonic}. Furthermore, \textit{Blade Runner 2049 }explores the technological feasibility and ethical considerations of 3D printing in the context of creating synthetic humans, a concept inching closer to reality due to advancements by firms such as Organovo \cite{neff2017printing}.

Researchers have explored the symbiotic relationship between science fiction and technological development, arguing that the genre can serve as a catalyst for innovation by providing a vision of potential future technologies \cite{michaud2017innovation}. Moreover, science fiction has been used as a tool for technology forecasting, helping to identify emerging trends and assess their potential impact on society \cite{firat2008technological}\cite{roberts2014evolving}.

\subsection{Video Games as a Medium for Future Technologies}
As the video game industry has evolved, it has increasingly served as a platform for exploring futuristic concepts and ideas. Video games offer a unique opportunity to engage with potential future technologies in an interactive manner, enabling players to immerse themselves in the game world and experience the consequences of these technologies first-hand \cite{shaffer2005video}\cite{smith2007disruptive}. Some recent studies have focused on the potential of video games for inspiring and predicting technological advancements, highlighting the role of game developers in creating plausible and engaging visions of the future \cite{lambeth2019developing}\cite{ pirker2020video}.

\subsection{Technologies in Cyberpunk 2077}
The specific technologies presented in Cyberpunk 2077 have also been the subject of academic inquiry, with researchers exploring the game's representation of various technological domains. Human augmentation, for instance, has been a subject of interest in various studies, examining the ethical, social, and philosophical implications of integrating technology into the human body \cite{bostrom2008ethical}\cite{Bostrom2009}\cite{warwick2010implications}. Similarly, brain-computer interfaces, as depicted in Cyberpunk 2077's \textit{braindance} concept, have been investigated in terms of their potential for revolutionizing human cognition and communication \cite{Lebedev2017}.

Other areas of research related to the game's themes include simulated reality \cite{kilteni2012sense}, digital representation and information access \cite{lee2021all}, smart homes and appliances \cite{cook2004smart}\cite{ gams2019artificial}, autonomous vehicles \cite{Thrun2010}, and advanced smart assistants \cite{knote2019classifying}. These studies provide insights into the current state of these technologies, their challenges, and potential future developments.

In summary, the literature highlights the potential of science fiction and video games, such as Cyberpunk 2077, in predicting and inspiring future technological advancements. By examining the game's portrayal of various technologies, this paper aims to contribute to the ongoing academic conversation about the potential trajectories of these advancements and their implications for society.


\section{Methodology}

In this study, we aim to systematically analyze the key technological themes and advancements presented in Cyberpunk 2077 and compare them with the current state of technology. To achieve this, we followed a three-step methodology, which we outline below.

\subsection{Data Collection}
First, we collected data from the Cyberpunk 2077 video game and its associated official materials, such as guides, trailers, and developer interviews. We focused on identifying the technological concepts and innovations present in the game, paying particular attention to the technologies that were emphasized in the presentation. This data served as our primary source for understanding the vision of the future as portrayed by the game developers.

\subsection{Thematic Analysis}
Next, we conducted a thematic analysis of the collected data to identify key themes and technological advancements within the game. We followed Braun and Clarke's six-phase approach to thematic analysis \cite{Braun2006}, which involved:

\begin{enumerate}
\item Familiarizing ourselves with the data by immersing ourselves in the game world and closely examining the identified technologies.
\item Generating initial codes by identifying interesting features of the data and systematically coding these features.
\item Searching for themes by collating the generated codes into potential themes.
\item Reviewing themes by refining the identified themes and ensuring that they are supported by the data.
\item Defining and naming themes by providing clear definitions and names for each theme.
\item Producing the final report by presenting the findings of the analysis in a coherent manner.
\end{enumerate}

This process allowed us to identify the key technological themes in Cyberpunk 2077 and organize them into coherent categories for further analysis.

\subsection{Comparison with Current Technologies}
Finally, we compared the identified themes and technological advancements with the current state of technology by reviewing the relevant literature and examining recent developments in the respective domains. This step involved assessing the current state of research, development, and implementation of the technologies in question, as well as identifying the primary challenges and opportunities for further advancement of the fields.

By following this methodology, we were able to systematically analyze the technological themes and advancements presented in Cyberpunk 2077 and evaluate their plausibility and relevance in the context of current technological progress.

\section{Themes and Technologies}

In this section, we present the key themes and technologies identified through our thematic analysis of Cyberpunk 2077. These themes encompass the core aspects of the game's portrayal of future technological advancements and provide a framework for understanding the game's vision of the future. A summary of the findings, including the links between depicted future technologies, current state of the art and possible research directions is shown in Table \ref{tab:summary}.

\begin{table*}[ht!]
\def\arraystretch{1.2}%
\setlength{\tabcolsep}{0.8em}
\label{tab:summary}
\centering
\caption{Main findings and ideas}
\begin{tabular}{|>{\centering\arraybackslash}m{3.2cm}|>{\centering\arraybackslash}m{4.3cm}|>{\centering\arraybackslash}m{4.3cm}|>{\centering\arraybackslash}m{4.3cm}|}
\hline
\textbf{\textit{Theme/Technology}} &\textbf{\textit{Future Cyberpunk 2077 Technology}} & \textbf{\textit{State of Current Technology}} & \textbf{\textit{Research Directions}} \\ 
\hline

Cybernetic Prosthetics & Advanced cyberimplants with augmented capabilities (Fig. \ref{fig:human_augmentation}) & Current implants, prosthetics, neuroprosthetics \cite{pilla2022cybernetic}\cite{ghadage2023reviewProstheticArms} & Improve hardware, software, signal processing \cite{bumbavsirevic2020BionicLimbsChallenges}; Advanced sensors, materials, security protocols \\ \hline

Artificial Biological Augmentations & All types of Lab-grown organs, novel tissues (Fig. \ref{fig:human_augmentation}) & Biologically engineered enhancements, tissue engineering \cite{neff2017printing} & Develop advanced biomaterials \cite{Parsons2021cellularTransplantation}\cite{Sohn2020OrganEngineering}; Overcome vascularization, innervation, immune compatibility \\ \hline

Genetic Modifications & Altered genetic code and modifications (Fig. \ref{fig:human_augmentation}) & CRISPR/Cas9, gene editing, genomics \cite{akram2023insightCRISPR} & Refine genome editing, accuracy \cite{akram2023insightCRISPR}; Address ethical, legal, societal implications \cite{joyner2019genetic}\cite{varillas2022genetics}\\ \hline

Brain-Computer Interfaces & Neuralware for seamless brain-device communication, (Fig. \ref{fig:BCI}) & EEG-based BCIs, invasive BCIs \cite{Sterniuk2021BCIsReview} & Enhance connectivity, miniaturization \cite{chandrasekaran2021BCIFutureDirections}; cybersecurity \cite{BROCAL2023BCIRisks}, novel environments \cite{qiu2023keyBCISpace}\\ \hline

Simulated Reality & Multi-sensory, fully realistic, immersive VR and AR systems (Fig. \ref{fig:BCI}) & VR, AR technologies, haptic devices \cite{zhan2020augmented}\cite{saredakis2020factors} &AI-generated content, realistic interactions \cite{dangxiao2019haptic}\cite{kugler2021state}; Haptic feedback, gesture recognition \\ \hline

Digital Twins & Virtual replicas of all objects and humans for enhanced control and simulation (Fig. \ref{fig:Digital_representation}) & Performance monitoring, predictive modeling \cite{vanderhorn2021digital} & Develop data exchange protocols, real-time processing \cite{de2020digital}; Integrate biometric, biological data \cite{de2020digital} \\ \hline

Advanced Biometric Authentication & Secure personalized access based on inmuttable physical characteristics (Fig. \ref{fig:Digital_representation}) & Fingerprint, facial recognition, iris scanning \cite{sarkar2020review} & Improve accuracy, security \cite{RYU2023BiometricAuthChallenges}; Develop continuous authentication, privacy techniques \\ \hline

Information Transfer and Security & Almost instantaneous, secure and wireless data exchange of large data (Fig. \ref{fig:Digital_representation}) & Modern wireless networks (5G/6G), encryption standards \cite{taleb20226g} & Develop communication protocols, encryption \cite{lovensemantic}; Explore semantic communication, edge computing \\ \hline

Smart Mirrors and Fashion Applications & Exhaustive analysis of appearance and ability to show modifications and fashion (Fig. \ref{fig:Smart_environments}) & Smart mirrors, augmented shopping \cite{Alboaneen2020ReviewSmartMirrors} & Improve computer vision, image recognition \cite{luce2019computer}; Real-time analysis, virtual clothing manipulation \cite{luce2019computer} \\ \hline

Smart Appliances and Home Automation & Autonomous, coordinated and personalized systems (Fig. \ref{fig:Smart_environments}) & Smart appliances, IoT integration \cite{sovacool2020smart} & Inter-device communication, machine learning \cite{dahlgren2021personalization}; Standards, protocols \cite{dahlgren2021personalization}; User behavior, intentions \cite{dahlgren2021personalization} \\ \hline

Autonomous Ground Vehicles & Fleets of driverless vehicles in urban settings with integrated traffic management (Fig. \ref{fig:Autonomous_vehicles}) & Testing self-driving cars, sensor fusion \cite{wang2020safety} & Improve SLAM, sensor fusion \cite{ahangar2021survey}; Develop V2V, V2I communication \\ \hline

Autonomous Flying Vehicles & Advanced drones and personal flying aircraft for urban mobility (Fig. \ref{fig:Autonomous_vehicles}) & Autonomous drones, quadcopters \cite{kakaletsis2021computer} & Enhance flight control, collision avoidance \cite{kurunathan2023machine}; Develop computer vision, AI algorithms \\ \hline

Autonomous Delivery Robots & Fleets of ground-based delivery robots for urban and remote areas (Fig. \ref{fig:Autonomous_vehicles}) & Limited deployment, urban deliveries \cite{Srinivas2022RobotDeliveries} & Navigation, obstacle detection \cite{hossain2023autonomous}; Dynamic routing, standardized communication \\ \hline

Self-aware AI & AI systems capable of complex reasoning, introspection, and self-improvement (Fig. \ref{fig:AI_assistants}) & Current AI, LLMs, neural networks \cite{marcus2022deephittingwallAGI} & Develop AI self-awareness, reasoning \cite{brcic2023impossibility}; Address ethical considerations, AI alignment, bias\\ \hline

AI-based Personal Assistants & Predictive AI assistants with adaptive learning and personalization (Fig. \ref{fig:AI_assistants}) & Current AI assistants, voice recognition \cite{hu2021can} & Emotion recognition \cite{wang2022iteratively}; Detect and interpret human emotions, predict user needs \cite{wei2022emergent} \\ \hline

\end{tabular}
\end{table*}

\subsection{Overarching Themes}

In addition to the specific technologies and advancements, Cyberpunk 2077 presents several overarching themes that pervade its vision of the future. Reflecting on the broader trends and principles that shape the development and application of new technologies in the game's world, these overarching themes provide a broader context for understanding the specific technologies and advancements present in Cyberpunk 2077, highlighting key principles and trends that shape the game's portrayal of the future.

\begin{enumerate}

\item\textbf{Ubiquitous Sensing:} \textit{(Everything Sensing)}

In the game's future, virtually every device, object, and even human beings are integrated with a multitude of sensors. This constant sensing enables real-time data collection and analysis in a truly distributed manner, which in turn drives more informed decision-making and responsive technologies.

\item\textbf{Interconnectivity: } \textit{(Everything Connected)}

The world of Cyberpunk 2077 is characterized by ubiquitous interconnectivity, with all devices, systems, and even humans constantly communicating and collaborating. This interconnectivity allows for the formation of very large-scale networks of sensors and data sources, enabling more efficient and coordinated responses to various challenges. 

\item\textbf{Autonomy:} \textit{(Everything Autonomous)}

Many technologies in the game are autonomous, meaning they can operate independently without direct human intervention. This autonomy extends to various devices and systems, ranging from computing resources and communication networks, to vehicles and robots.

\item\textbf{Multimodality:} \textit{(Everything Multimodal)}

Technologies in Cyberpunk 2077 often embrace multimodality, capturing and processing various types of sensory data simultaneously. This multimodal approach enables more comprehensive and nuanced understanding and analysis of complex situations.

\item\textbf{Personalization:} \textit{(Everything Personalized)}

A strong emphasis is placed on  the personalization of the game's world, with technologies adapting to individual users' locations, needs, and preferences. This personalization allows for more tailored and effective solutions to various challenges and enhances overall user experiences, allowing people to effectively experience the world in a truly unique manner.

\item\textbf{Bio-Integration:}  \textit{(Everything Bio-Integrated)}

The game's vision of the future also includes the convergence of the physical, artificial, and biological worlds. Technologies are seamlessly integrated into human bodies and natural environments, creating new hybrid systems that combine the strengths of each domain.

\item\textbf{Digital Security:}  \textit{(Everything Secure)}

The world of Cyberpunk 2077 shows unwavering dedication to securing all digital interactions, from personal communication to vital infrastructure. In this hyper-connected society, where everything is linked digitally, constant readiness against cyber threats is essential. The narrative highlights the need for robust safeguards and vigilance against potential breaches.

\end{enumerate}

The main technologies of the game have been grouped in six different thematic groups: Human Augmentation, Brain Computer Interfaces, Digital Representation and Information Access, Smart Environments and Personalization, Autonomous Vehicles and Transportations and Self-aware AI and Assistants. 

\subsection{Human Augmentation}
\begin{figure}[ht!]
  \begin{center}
    \includegraphics*[width=0.48\textwidth]{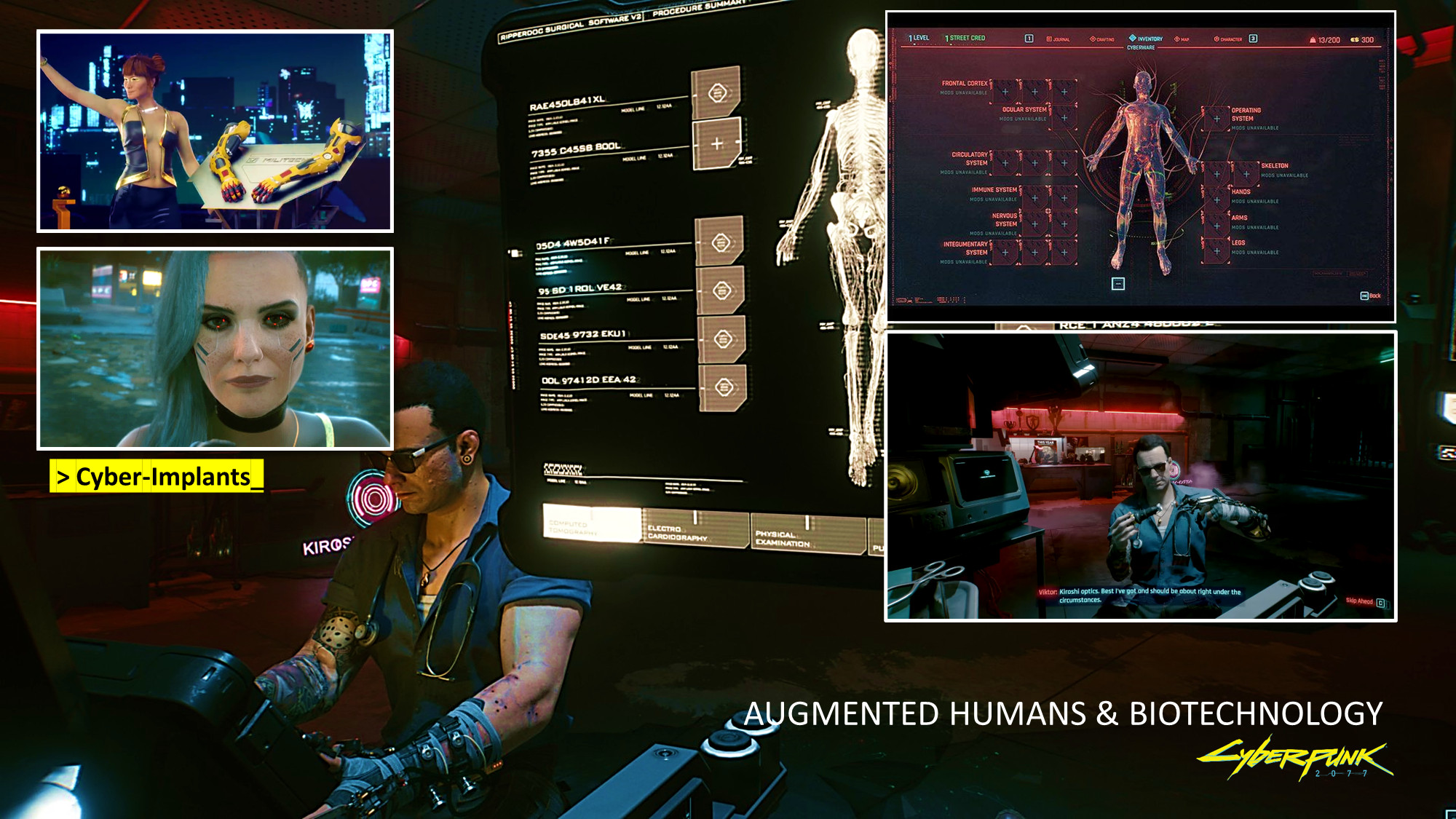}
  \end{center}
  \vspace{-3mm}
    \caption{Human Augmentation technology as depicted in Cyberpunk 2077.}
  \label{fig:human_augmentation}
  \vspace{-3mm}
\end{figure}

Cyberpunk 2077 envisions a future where humans are augmented through biotechnology, cybernetic implants, and genetic modifications. However, in this envisioned society, not all individuals opt for or can afford these enhancements, revealing disparities and philosophical tensions among its denizens. This theme includes technologies such as cybernetic prosthetics, artificial biological augmentations, and genetic enhancements that expand human capabilities beyond their natural limits. Fig \ref{fig:human_augmentation} shows examples of human augmentation technologies as depicted in Cyberpunk 2077.

\textbf{Cybernetic Prosthetics} replace or enhance body parts with advanced mechanical or electronic components, providing users with improved strength, agility, or sensory perception \cite{pilla2022cybernetic}. In Cyberpunk 20177, some examples include robotic limbs, advanced ocular implants, and neural interfaces. Today, prosthetic limbs have evolved significantly, incorporating advanced materials, sensors, and electronics to provide improved functionality. Some prosthetics can be controlled by the user's muscle signals or through brain-computer interfaces, while others offer sensory feedback \cite{pilla2022cybernetic}\cite{ghadage2023reviewProstheticArms}. However, these prosthetics are still limited in terms of sensory perception, dexterity, and overall performance compared to the advanced prosthetics portrayed in Cyberpunk 2077. To reach the level of cybernetic prosthetics in the game, research should focus on developing more sophisticated hardware, software, and signal processing techniques that can better interpret user intent and accurately mimic natural movements \cite{bumbavsirevic2020BionicLimbsChallenges}. Additionally, integrating advanced sensors, materials, and power systems will be crucial for creating more realistic and high-performance prosthetics \cite{liang2019high}\cite{wolf2020advanced}. As the trust users place in their cybernetic enhancements hinges on the guarantee that they are secure from external manipulations, future research should also prioritize establishing robust security protocols and standards.

\textbf{Artificial Biological Augmentations} involve the incorporation of synthetic or lab-grown biological components into the human body, such as replacement organs, synthetic muscles, or artificial blood. These augmentations can improve health, vitality, or physical performance. In the present time, advances in regenerative medicine and tissue engineering have led to the development of lab-grown organs and tissues, such as skin, blood vessels, and cartilage \cite{neff2017printing}. Currently, the ethical implications of using human augmentation are still carefully considered, ensuring they are used for genuine therapeutic purposes and not merely for enhancing one's natural abilities beyond ethical boundaries \cite{Parsons2021cellularTransplantation}. However, these technologies are still in the early stages of development, and widespread clinical application remains limited. To achieve the level of artificial biological augmentations seen in Cyberpunk 2077, researchers need to continue exploring methods for generating functional, lab-grown tissues and organs that can seamlessly integrate with the human body. This includes the development of advanced biomaterials, cell sources, and fabrication techniques, as well as overcoming challenges related to vascularization, innervation, and immune compatibility \cite{Parsons2021cellularTransplantation}\cite{Sohn2020OrganEngineering}.

\textbf{Genetic Modifications} in Cyberpunk 2077 entail the alteration of an individual's genetic code to enhance physical or mental traits, such as increased intelligence, physical strength, or resistance to diseases. To date, genome editing technologies such as CRISPR/Cas9 have facilitated gene modification with remarkable precision, thereby contributing to significant advancements in both gene therapy and the creation of genetically modified organisms \cite{akram2023insightCRISPR}. Despite these advancements, a comprehensive understanding of the genetic influences on human traits and diseases remains elusive \cite{mcguire2020roadGnomics}. Recent scholarly reviews indicate that the complexities of genetic impact on specific attributes, such as sports performance, have yet to be fully elucidated, suggesting that similar limitations may apply to the enhancement of other human traits \cite{joyner2019genetic}\cite{varillas2022genetics}. To achieve the level of genetic modifications depicted in the game, researchers must continue to refine genome editing techniques to increase their accuracy, efficiency, and safety. This includes developing new gene editing tools, delivery methods, and strategies for minimizing off-target effects. Additionally, addressing ethical, legal, and societal implications of human enhancement through genetic modifications will be crucial for their responsible implementation.

\subsection{Brain-Computer Interfaces and Simulated Reality}

\begin{figure}[ht!]
  \begin{center}
    \includegraphics*[width=0.48\textwidth]{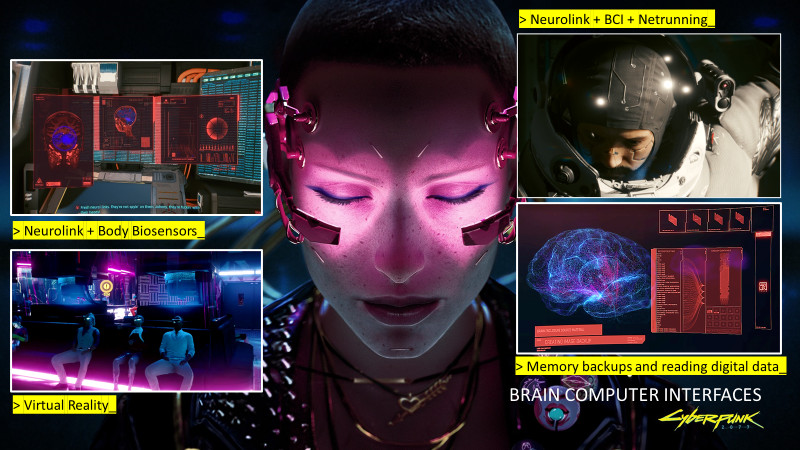}
  \end{center}
  \vspace{-3mm}
    \caption{Brain Computer Interfaces as depicted in Cyberpunk 2077.}
  \label{fig:BCI}
\end{figure}

The game introduces the concept of \textit{braindance}, an advanced recording technology capable of capturing and replaying sensory input and external sensor data. This innovation enables users to immerse themselves in past experiences on various sensory levels. This theme also encompasses significant advancements in simulated reality, facilitating the creation of exceptionally realistic virtual environments suitable for diverse purposes, including training, education, and entertainment. Within the Cyberpunk 2077 universe, certain inhabitants are deeply immersed in this hyper-connected environment, often prioritizing their experiences through such interfaces. However, this heightened reliance on technology has given rise to mental health concerns, notably the condition known as cyberpsychosis. Figure \ref{fig:BCI} provides visual examples of Brain-Computer Interface (BCI) technology as portrayed in Cyberpunk 2077.

\textbf{Brain-Computer Interfaces} (BCIs) play a critical role in Cyberpunk 20277, depicted as devices facilitating direct communication between the human brain and external devices, such as neuralware and \textit{braindance} systems. Modern BCIs have made significant progress in recent years, with applications ranging from assisting individuals with disabilities to controlling virtual environments \cite{Zabcikova2022BCIRecentAdvances}. However, current BCIs are primarily non-invasive and have limitations in terms of signal quality, bandwidth, and user experience compared to the highly advanced BCIs depicted in the game \cite{Sterniuk2021BCIsReview}. Moreover, as these interfaces access and transmit neural data, concerns about data security and unauthorized access emerge, and need cybersecurity measures to safeguard users' neural and personal information \cite{BROCAL2023BCIRisks}. To advance the state of brain-computer interfaces (BCIs) to the level depicted in Cyberpunk 2077, targeted research is needed to improve their connectivity, miniaturization, and biocompatibility. This involves work on minimally-invasive methods, faster data transmission, new materials, improved signal storage, and better algorithms for signal processing \cite{chandrasekaran2021BCIFutureDirections}. Such advancements in BCIs are also considered crucial for their application in future manned space missions \cite{qiu2023keyBCISpace}, an aspect highlighted in Cyberpunk 2077.


\textbf{Simulated Reality} in Cyberpunk 2077 features highly immersive simulated reality experiences, including virtual reality (VR) and augmented reality (AR) systems that can seamlessly blend digital content with the physical world. VR and AR technologies have seen rapid advancements in recent years, with applications in gaming, education, and training\cite{rojas2023systematic}\cite{xie2021review}. However, these systems still have limitations in terms of display resolution, field of view, induction of motion sickness, and user interaction compared to the near-perfect simulated reality portrayed in Cyberpunk 2077\cite{zhan2020augmented}\cite{saredakis2020factors}. However, to reach the level of simulated reality seen in the game, research should focus on developing AI-generated, procedural content, improving display technologies, and creating realistic interactions within virtual environments. This includes advancing the development of haptic feedback, gesture recognition, and environmental sensing technologies, as well as optimizing software and hardware for more immersive experiences\cite{dangxiao2019haptic}\cite{kugler2021state}.


\subsection{Digital Representation and Information Access}
\begin{figure}[ht!]
  \begin{center}
    \includegraphics*[width=0.48\textwidth]{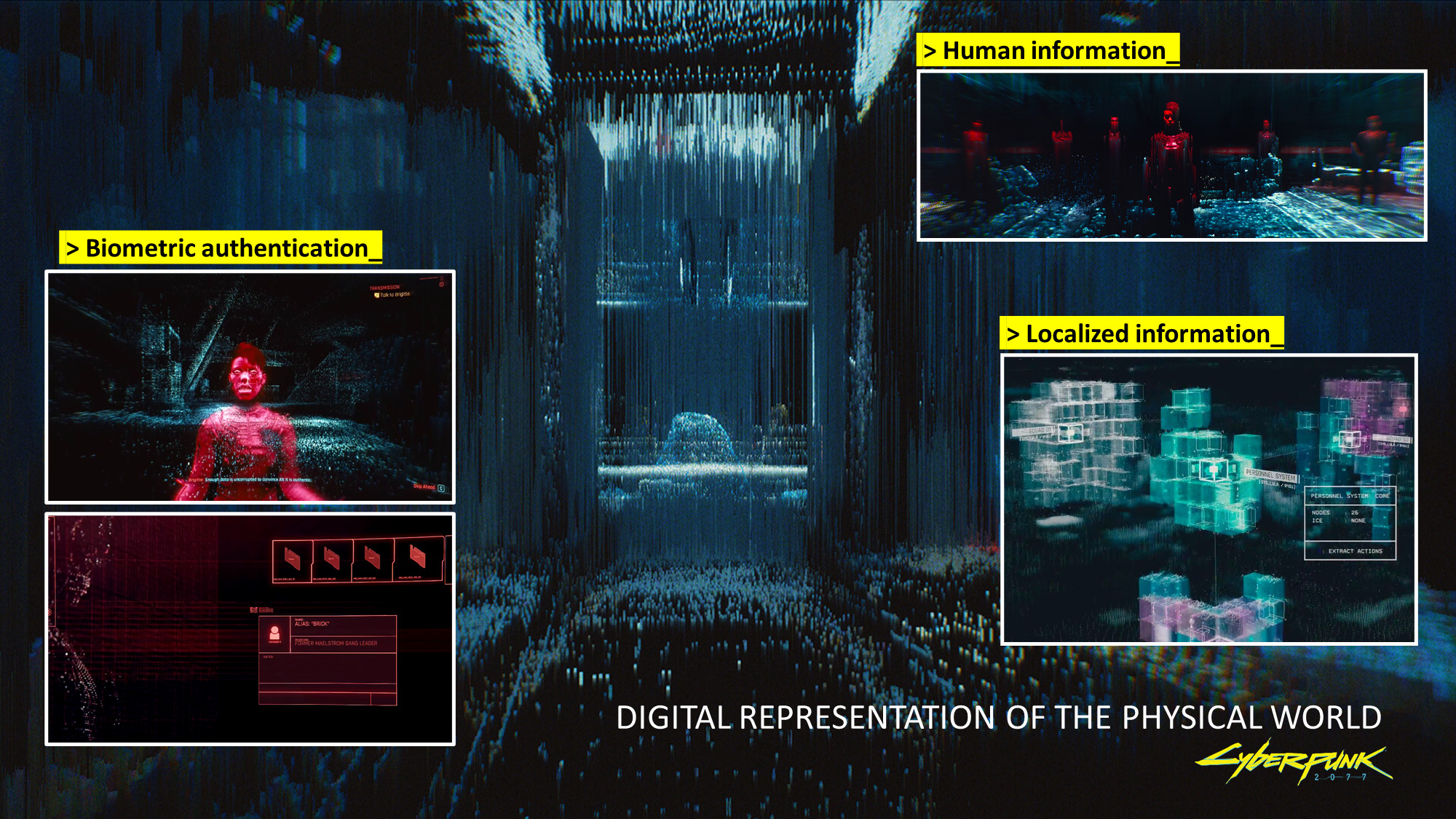}
  \end{center}
  \vspace{-3mm}
    \caption{Digital representation of the world as depicted in Cyberpunk 2077.}
  \label{fig:Digital_representation}
\end{figure}

Cyberpunk 2077 presents a future where digital information is seamlessly integrated into the physical world through standardized protocols and real-time authentication based on biometrics. Amidst these advancements, unlawful access to information emerges as the most prevalent act of theft, highlighting the perpetual game between security and intrusion. This theme includes the development of digital twins, advanced biometric access systems, and improved information transfer and security. Fig \ref{fig:Digital_representation} shows examples of the Cyberspace, the Digital Representation of the world as depicted in Cyberpunk 2077.

\textbf{Digital Twins} are envisioned in the world of Cyberpunk 2077 in a way where virtual replicas of physical objects or systems are commonplace, and show ubiquitous, and instantaneous capabilities. They allow for enhanced simulation, monitoring, and optimization of real-world assets. Digital twin technology is already being used in various industries, such as manufacturing, infrastructure, and aerospace, for performance monitoring, predictive maintenance, and decision-making support\cite{vanderhorn2021digital}. However, the level of integration and sophistication depicted in Cyberpunk 2077 is still beyond the capabilities of current technology. To achieve this level of digital twin integration, research should focus on developing standardized protocols for data exchange, improving real-time data processing, and enhancing the scalability and interoperability of digital twin systems. This includes the development of advanced machine learning and AI techniques for data analysis, optimization, and prediction. In addition, research is needed into integrating biometric and biological data into \textit{human digital twins} that can populate the virtual replicas of the physical world\cite{de2020digital}.

\textbf{Advanced Biometric Authentication} systems provide the Cyberpunk 2077 works with secure, personalized access to various services and facilities based on unique physical characteristics, such as fingerprints, iris patterns, or facial features. Currently, biometric authentication is widely used in modern security systems and personal devices, with technologies such as fingerprint scanners, facial recognition, and voice recognition becoming increasingly common\cite{sarkar2020review}. However, the game presents a future where biometrics are globalized, more pervasive and integrated into everyday life. To reach the level of biometric authentication seen in Cyberpunk 2077, research should focus on improving the accuracy, reliability, and security of biometric systems, especially when facing adversarial attacks. In addition, future research efforts should prioritize the development of continuous authentication methodologies, ensuring that users are continuously verified throughout their interactions with systems, enhancing both security and user experience \cite{RYU2023BiometricAuthChallenges}. This includes developing new sensors, algorithms, and privacy-preserving techniques to ensure the robustness and user-friendliness of the solutions.

\textbf{Information Transfer and Security} mechanisms, allowing for rapid and secure exchange of data across various networks and devices of the Cyperpunk 2077 universe. In recent times, modern wireless communication networks (5G/6G) have made significant strides in terms of speed, reliability, and security, and future architectures plan to increase extend their capabilities and flexibility beyond what is usable now\cite{taleb20226g}. However, they are not yet seamlessly integrated and foolproof. To achieve the level of information transfer and security presented in the game, research should focus on developing new communication protocols, encryption algorithms, and network architectures that can support the rapid, secure, and seamless exchange of data. This includes the exploration of semantic communication, edge computing, and advanced cybersecurity techniques to protect against potential threats and vulnerabilities\cite{lovensemantic}.

\subsection{Smart Environments and Personalization}
\begin{figure}[ht!]
  \begin{center}
    \includegraphics*[width=0.48\textwidth]{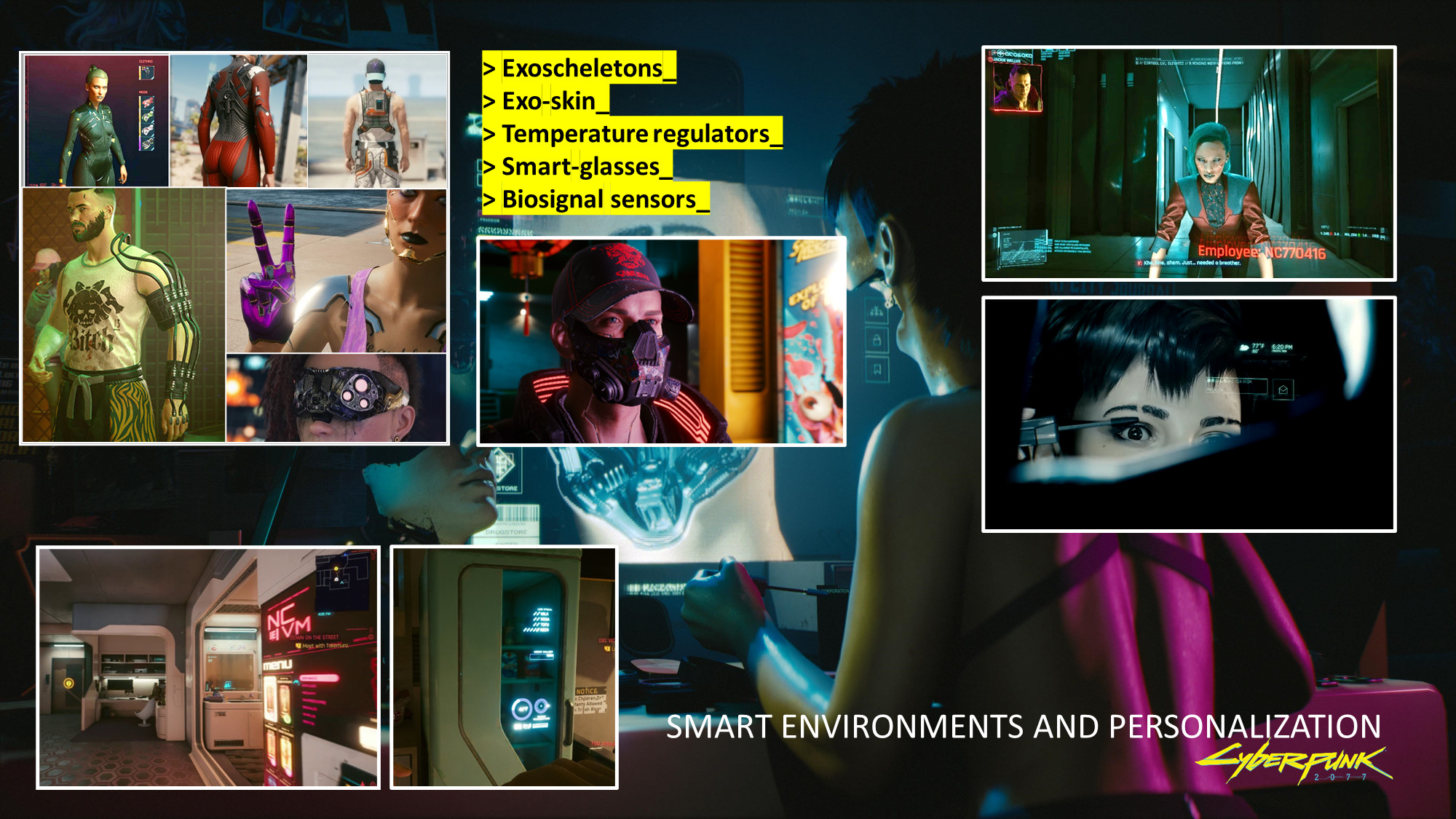}
  \end{center}
  \vspace{-3mm}
    \caption{Personalized Smart environments in Cyberpunk 2077.}
  \label{fig:Smart_environments}
\end{figure}

The game envisions a future where homes and appliances are smart, interconnected, and highly personalized. Technologies in this theme include different home appliances such as smart mirrors, fashion applications, smart fridges, and various home automation systems that adapt to individuals' needs and preferences. Fig \ref{fig:Smart_environments} shows examples of different smart environments and personalization technologies present in Cyberpunk 2077.

\textbf{Smart Mirrors and Fashion Applications} can analyze a user's appearance and suggest fashion choices or cosmetic adjustments, enabling users to change their clothing or appearance virtually. Even nowadays, smart mirrors with limited capabilities, such as displaying weather or news updates, already exist \cite{Alboaneen2020ReviewSmartMirrors}. Current technology is progressing to offer detailed fashion advice or enable virtual clothing changes, but challenges remain in how to input commands in real-time with great accuracy\cite{ogunjimi2021smart}. To achieve the level of smart mirrors and fashion applications seen in the game, research should focus on improving computer vision, image recognition, and AR/VR technology. This includes developing algorithms for real-time clothing and appearance analysis, virtual clothing manipulation, but specially, seamless AR/VR integration\cite{luce2019computer}.

\textbf{Smart Appliances and Home Automation} systems that coordinate their behavior, learn from users' preferences, and adapt to individual needs are omnipresent in Cyberpunk 2077. Nowadays, smart appliances and home automation systems are already available, offering a range of capabilities such as remote control, energy management, and simple learning algorithms. However, current systems generally have limited coordination and personalization capabilities compared to those portrayed in the game\cite{sovacool2020smart}. To achieve the level of smart appliances and home automation seen in Cyberpunk 2077, research should focus on improving inter-device communication, machine learning algorithms, and user behavior analysis. This includes developing standards and protocols for seamless device integration, advanced algorithms for learning user preferences and predicting needs, and improved techniques for analyzing user behavior and intentions. Research into coordination and personalization of devices with a complex socio-material perspective is paramount\cite{dahlgren2021personalization}.

\subsection{Autonomous Vehicles and Transportation}
\begin{figure}[ht!]
  \begin{center}
    \includegraphics*[width=0.48\textwidth]{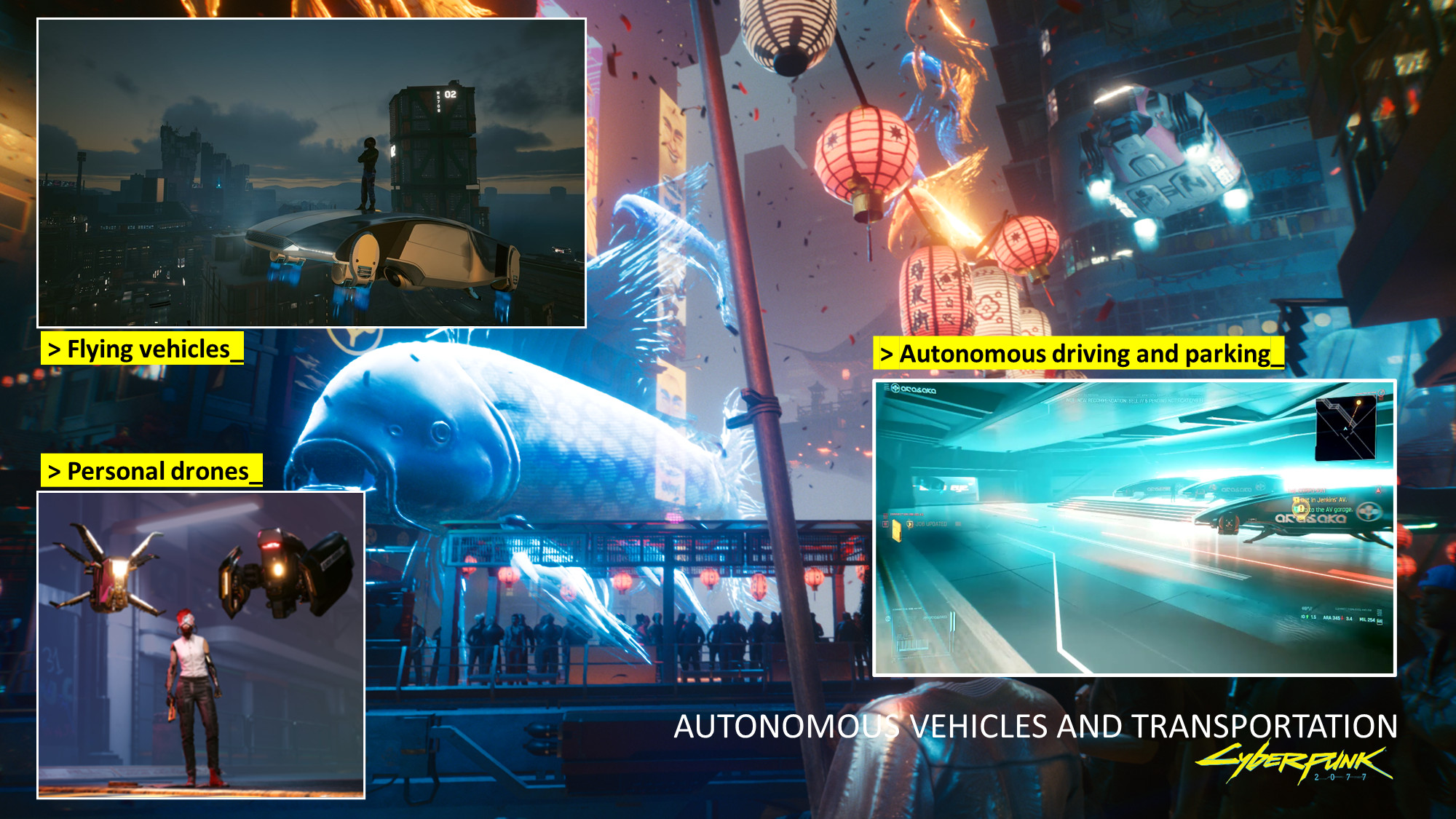}
  \end{center}
   \vspace{-3mm}
    \caption{Autonomous vehicles and transportation in Cyberpunk 2077.}
  \label{fig:Autonomous_vehicles}
\end{figure}

Fully autonomous vehicles, including cars and drones, are a key feature of Cyberpunk 2077's futuristic setting, where they often coexist with regular ones. These smart vehicles, rely on advanced sensor and communication systems to navigate complex urban environments and interact with unpredictable elements, such as pedestrians and other vehicles. Fig \ref{fig:Autonomous_vehicles} shows flying cars, personal drones and autonomous driving and parking technology as seen in Cyberpunk 2077.

\textbf{Autonomous Ground Vehicles} are showcased in the game as driverless cars and taxis capable of traversing intricate urban settings with minimal human intervention, in a coordinated manner. In the current state of the art, autonomous ground vehicles have made significant progress in recent years, with companies such as (e.g.) Tesla, Waymo, and Cruise developing and testing self-driving cars. However, current autonomous vehicles still require human oversight and are not as advanced and fully autonomous as those depicted in the game\cite{wang2020safety}. To reach the level of autonomy portrayed in Cyberpunk 2077, research should focus on improving simultaneous localization and mapping (SLAM), sensor fusion, and decision-making algorithms for vehicles operating in dynamic and complex environments. Additionally, the development of standardized communication protocols and infrastructure to support vehicle-to-vehicle (V2V) and vehicle-to-infrastructure (V2I) communication will be crucial in facilitating safe and efficient autonomous transportation. Improvements of self-driving models to handle out-of-distribution examples and fringe cases need to be greatly improved\cite{ahangar2021survey}.

\textbf{Autonomous Flying Vehicles} such as drones and personal aircraft, are used in the Cyberpunk 2077 universe for transportation, surveillance, and delivery purposes. While autonomous drones are currently available for various applications, including aerial photography, mapping, and inspection, fully autonomous personal aircraft are still in the early stages of development and testing. To achieve the level of autonomous flying vehicles seen in the game, research should focus on enhancing flight control systems, collision avoidance, and air traffic management for operation in urban environments. This includes developing advanced computer vision for sensing and understanding\cite{kakaletsis2021computer}, AI algorithms for path planning, real-time decision-making, and robust communication systems to facilitate seamless interaction with other aerial vehicles and ground-based infrastructure\cite{kurunathan2023machine}.

\textbf{Autonomous Delivery Robots} in Cyberpunk 2077 are small ground-based delivery robots that navigate sidewalks and streets to transport packages and goods. Companies such as Starship Technologies, Nuro, and Amazon have developed and tested autonomous delivery robots, but their deployment is limited and not yet widespread and ubiquitous \cite{Srinivas2022RobotDeliveries}. To achieve the level of autonomous delivery robots seen in the game, research should focus on improving navigation algorithms, obstacle detection and avoidance, and human-robot interaction to ensure the safe and efficient operation of these robots in complex urban environments. Autonomous robot-driven delivery systems should address realistic constraints, incorporate uncertainty through stochastic modeling, and enable dynamic routing through multi-objective optimization. Additionally, the development of standardized communication protocols and infrastructure to support robot-to-robot (R2R) and robot-to-infrastructure (R2I) communication will be vital in facilitating their large-scale deployment\cite{hossain2023autonomous}.

\subsection{Advanced Artificial Intelligence and Smart Assistants}
\begin{figure}[ht!]
  \begin{center}
    \includegraphics*[width=0.48\textwidth]{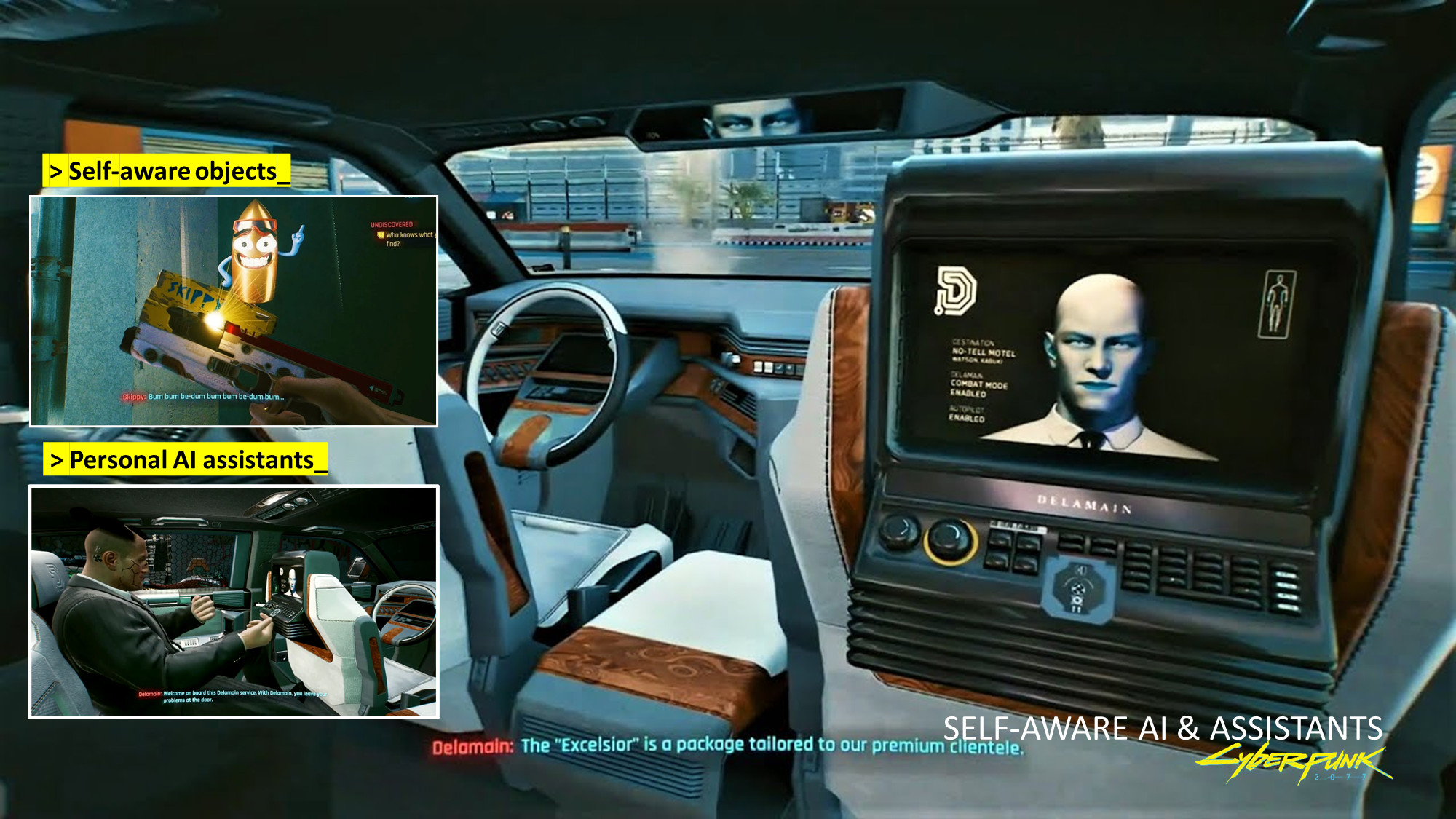}
  \end{center}
  \vspace{-3mm}
    \caption{Self-aware Artificial Intelligence as shown in Cyberpunk 2077.}
  \label{fig:AI_assistants}
\end{figure}

The game showcases advanced AI systems that are self-aware and capable of personalized interactions based on users' traits, emotions, and intentions. These smart assistants provide users with context-aware support and adapt their behavior to better serve individual needs.  Fig \ref{fig:AI_assistants} shows flying cars, personal drones and autonomous driving and parking technology as seen in Cyberpunk 2077.

\textbf{Self-aware AI} systems are portrayed in Cyberpunk 2077as being capable of complex reasoning, decision-making, and natural interactions with humans. While AI has made significant progress in recent years, current AI systems are generally narrow in their capabilities and are not yet fully self-aware or capable of general intelligence \cite{marcus2022deephittingwallAGI}. In the very recent times, Large Language Models (LLM) have shown incredible capabilities beyond what was previously possible, and the rapid pace of development is only expected to bring new possibilities at a fast pace. However, in order to achieve the level of self-aware AI seen in the game, research should focus on developing algorithms and architectures that enable AI systems to reason about their own existence, understand their limitations, and adapt to new situations. This includes advancing areas such as artificial general intelligence (AGI), machine learning, and cognitive computing, while taking enough time to discuss ethical topics that include AI alignment, and bias awareness\cite{brcic2023impossibility}.

\textbf{AI-based Personal Assistants} are featured in Cyberpunk as agents that are able to predict user needs, recognize emotions, and adapt their behavior accordingly. Currently, AI-based personal assistants like Siri, Alexa, and Google Assistant have become increasingly popular, but their capabilities are limited compared to those portrayed in the game. While the most widespread smart assistants used to rely on predefined responses and were not able to respond to a wide variety of scenarios in a credible manner, recent advances in LLMs, including fine-tuning of the models for human conversation and Reinforcement Learning from Human Feedback, have pushed the field to new heights, increasing the usability of the systems almost exponentially. However, current assistants, although able to influence human cognition, are still not able to fully understand or adapt to users' emotions and intentions\cite{hu2021can}. To achieve the level of AI-based personal assistants seen in the game, research should focus on improving not only natural language processing, but emotion recognition, and context-aware reasoning for chain of thougths\cite{wang2022iteratively}. This includes developing advanced algorithms for understanding and generating natural language from multimodal data\cite{driess2023palm}, but most importantly detecting and interpreting human emotions, and predicting user needs based on context and history\cite{wei2022emergent}.


\section{Discussion}
Our analysis of Cyberpunk 2077's vision of the future reveals a range of themes and technologies with the potential to significantly impact society, the economy, and culture. While some of these technologies are already emerging or in development, many remain speculative or face significant challenges to achieve the level of sophistication portrayed in the game.

One notable implication of the widespread adoption of human augmentation technologies is the potential for exacerbating existing social inequalities\cite{lutz2019digital}. Access to enhancements may be limited by factors such as cost, availability, and social acceptance, potentially leading to a divide between augmented and non-augmented individuals. This divide may manifest in terms of employment opportunities, social mobility, and even basic human rights. 

The rise of brain-computer interfaces and immersive simulated realities introduces new considerations for mental health\cite{kaimara2022could}. Prolonged immersion might blur the lines between real and virtual experiences, leading to cognitive overload, addiction, or challenges in distinguishing reality from the virtual. Educational systems and cognitive development patterns might also evolve in unforeseen ways as instant information access becomes commonplace.

As AI approaches higher levels of self-awareness and personalized interaction, ethical considerations about AI consciousness emerge\cite{schneider2019artificial}. These self-aware AIs could be treated as mere tools or as entities deserving of rights and recognition. Furthermore, the challenge of AI alignment — ensuring that the AI's goals and behaviors align with human intentions and values — becomes increasingly crucial\cite{alignment}. As AI systems become more complex and autonomous, there's a rising urgency to ensure they act in ways beneficial to humanity and avoid unintended consequences.  This choice has profound implications for AI-human relationships, leading to shifts in societal norms and the potential for new dynamics in how humans relate to technology.

The increasing interconnectivity and autonomy of technologies also raise concerns about privacy, security, and control. As devices and systems become more capable of sharing and processing data, the potential for misuse or abuse of this information grows\cite{christianto2018we}. Ensuring robust security and privacy protections will be crucial in maintaining trust and minimizing the risks associated with these advancements.

Environmental concerns also arise, particularly when contemplating the sustainability of producing and maintaining these high-tech devices\cite{riekki2021research}\cite{ropke2012unsustainable}. Sustainable practices in both production and disposal phases of augmentation technologies become paramount to mitigate environmental impacts.

The focus on personalization and bio-integration in Cyberpunk 2077's world highlights the need for ethical considerations in the development and application of new technologies\cite{yeung2018five}. Balancing the benefits of customization and individualized solutions with concerns about surveillance, manipulation, and loss of autonomy will be an ongoing challenge\cite{karwatzki2017beyond}.

A key insight from our analysis is the importance of designing technologies that can coexist with existing preferences and systems. Despite the rapid pace of technological change, it is important to recognize that not all individuals, organizations or societies will adopt or embrace new innovations uniformly\cite{hertzum2021technology}. Designing technologies that can accommodate a range of preferences, abilities, and cultural contexts will help ensure more equitable and inclusive outcomes\cite{knez2023technology}.

\section{Conclusion}
In this paper, we have explored the themes and technologies presented in the video game Cyberpunk 2077 as a lens through which to envision potential future technological advancements. Our analysis demonstrates the value of science fiction and video games as tools for stimulating imagination and fostering critical discussion about the direction and implications of emerging technologies.

While the game's portrayal of the future is undoubtedly speculative and stylized, it offers valuable insights into the opportunities and challenges associated with various technological advancements. By examining these themes and technologies, we can better anticipate the potential impact of new innovations on society, the economy, and culture, and inform the development of technologies that are more inclusive, ethical, and adaptable to a diverse range of needs and preferences.

Ultimately, the study of science fiction and video games like Cyberpunk 2077 can serve as a catalyst for interdisciplinary dialogue, collaboration, and exploration, helping to shape a more informed and responsible approach to the development and application of future technologies.


\bibliographystyle{ieeetr}
\bibliography{Cyberpunk.bib}

\end{document}